\documentclass[12pt]{article}
\usepackage{amssymb,amsmath,epsfig}
\begin{document}

\Large\textbf {Application of phenomenological Landau approach to the description of magnetic phase transitions in the ferromagnetic superconductor at ambient pressure}\\

\medskip

\normalsize
\textbf{Diana V. Shopova, sho@issp.bas.bg}
\medskip

\normalsize
Institute of Solid State Physics, Bulgarian Academy of Sciences, \\
1784 Sofia, Bulgaria


\begin{abstract}
We study the possibility to apply phenomenological approach to the description of magnetic transitions in UGe$_2$ with the help of Landau free energy expanded to $8^{th}$ order in magnetisation. The analysis shows that for certain values of parameters in front of $M^4$, and $M^6$ terms in the free energy there id possibility for the appearance of two successive phase transitions between the low-magnetisation and high-magnetisation phase with the same structure. We establish the relation of  the parameters in Landau energy, for which the phase transition from the disordered to low-magnetisation phase is of second order as the experimental data shows. Within our approximation the transition between two magnetically ordered phases is of first order.
\end{abstract}

\begin{figure}[!ht]
\begin{center}
\includegraphics[scale=0.55]{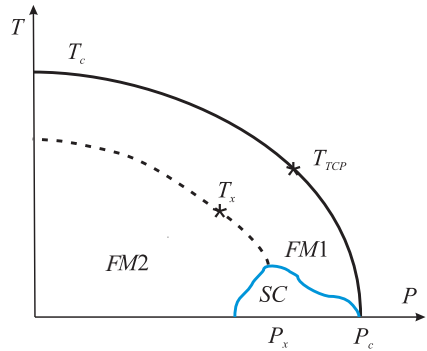}
\end{center}
\caption{The schematic temperature-pressure phase diagram of UGe$_2$. FM1 and FM2 describe the low- and high-magnetisation phases, respectively. The blue curve and the area below it describe the superconducting phase.}
\label{Fig1}
\end{figure}

\textbf{1. Introduction}\\
The most unique feature of uranium $5 f$-systems is the coexistence of the superconductivity and ferromagnetism carried by the same $5f$-electrons in
UGe2, URhGe, and UCoGe~\cite{Pfleiderer:2009},~\cite{Saxena:2000},~\cite{Gasparini:2007}. All the known materials where ferromagnetic order coexists microscopically with superconductivity are uranium compounds with substantially reduced magnetic moments compared to the free ion values. The magnetic moment of uranium compounds varies from nearly free ion values as for UGe$_2$, for which the spontaneous magnetic moment, experimentally established is  1.41$ \mu_B/$U ~\cite{Tateiwa:2014} to  very small values as for UCoGe - 0.039 $\mu_B/$U- itinerant case. \\
 UGe$_2$ is the first discovered ferromagnetic superconductor, where superconductivity appears only under pressure deep inside the ferromagnetic phase. It has orthorhombic crystal structure, where the U-atoms form coupled zigzag chains along the a-axis. The magnetic moments are aligned along the direction of the chains. In~Fig. (\ref{Fig1}) the schematic pressure-temperature phase diagram for UGe$_2$ is shown. The paper~\cite{Tateiwa:2018} demonstrates experimental phase diagram that summarises new experimental data and details. \\
 The transition at ambient pressure from paramagnetic to weakly polarized ferromagnetic phase (FM1) occurs at T$_c$ = 52.6 K  through second order phase transition with  M$_0 = 0.9 \mu_B$/U. With the increase of pressure, this transition changes from second to first order at trictical point T$_{TCP} \simeq 22 $ K, and  $P_{TCP} \approx 1.42$ GPa, both ferromagnetism and superconductivity  disappear at pressure $\sim 1.5$ GPa and T=0.\\
When  temperature is lowered a second ferromagnetic phase (FM2) appears at ambient pressure with the same magnetic structure and stronger magnetic polarization -1.41$\mu_B/$U ~\cite{Tateiwa:2014} at T$_x\approx$ 30 K; (see ~Fig. (\ref{Fig1})). Superconductivity is found in a limited pressure range between 1.0 and 1.5 GPa with a maximum transition temperature $\approx 0.7$ K at p$_x$ around the transition form FM1 to FM2 phase. There, the magnetic  moment is $\approx 1 \mu_B/U$ and the transition between two ferromagnetic phases is of first order at the critical end point (CEP: T$_{CEP} \sim 7 $ K, P$_{CEP} \sim 1.16 $ GPa).\\
It is experimentally established that UGe$_2$ has very strong uniaxial magnetic anisotropy~\cite{Onuki:1992}, and at some earlier stages of study, it has been considered that 3D Ising model can well describe its magnetic properties -  see, for example,~\cite{Shopova:2009}, and for recent review~\cite{Aoki:2019}. Experimental studies on the critical behaviour of UGe$_2$  within the critical regions of magnetic phase transitions~\cite{Tateiwa:2014} which are also theoretically  justified show that the  critical exponents for UGe$_2$ do not belong to any known universality class including 3D Ising model and anisotropic next nearest neighbor 3D Ising model. One of the reasons for this lays in the dualism of 5f-electrons in UGe$_2$ where two subsets of itinerant and localised 5f electrons exist. This fact is experimentally established by positive muon relaxation measurements where the  itinerant magnetic moment at ambient pressure is found to be $\approx $ 0.02$ \mu_B$ ~\cite{Yaouanc:2002}. This dualism of 5f-electrons is confirmed by magnetic, transport and specific heat experiments, see, for example, ~\cite{Troc:2012}. Understanding the magnetic phase transitions from microscopic point of view is important for understanding the appearance of superconductivity - for review, see for example ~\cite{Huxley:2015}, but there is no up to now consensus on this problem.\\
Usually microscopic models are focused on the phase transition between FM1 and FM2 around the critical end point, where superconducting phase appears under pressure ~\cite{Fidrysiak:2019} although at ambient pressure this problem is not yet resolved, especially the order of transition between FM1 and FM2. In earlier studies, it is proposed that the electronic density of states has two very closely placed peaks near the low-dimensional anisotropic Fermi surface, whose topology  changes by the appearance of magnetisation ~\cite{Sandeman:2002}. This calculation is performed at T=0 with the help of Stoner theory for itinerant magnetism. \\
Recently, an estimate is made on the basis of Takahashi's spin fluctuation theory about the itinerancy of 5f-electron systems in actinides ~\cite{Tateiwa:2017} and the comparison of this phenomenological approach with the experimental data shows that UGe$_2$ within the criterion used in the paper is an intermediate case between strongly localised and itinerant 5f-electron system.\\
In this paper we will consider the ferromagnetic phase transitions in UGe$_2$ at ambient pressure in the absence of external magnetic field. There are few experimental measurements in this  regime, and some authors claim that there is not a real phase transition between FM1 and FM2, but only a crossover~\cite{Fidrysiak:2019}. The positive muon spin rotation measurements of ~\cite{Sakarya:2010}, show unambiguously that at ambient and low pressure a real thermodynamic transition occurs between FM1 and FM2 phases.\\
We will apply to the description of the magnetic phase transitions in UGe$_2$ at ambient pressure the general phenomenological Landau approach by expanding the Landau free energy to $8^{th}$ order in magnetisation. From the general theory \cite{Izyumov:1984}; \cite{Toledano:1987}, the inclusion of M$^8$ leads to the appearance of two ordered  isostructural phases depending on the sign and magnitude of coefficients in front of M$^4$ and M$^6$ terms in the Landau free energy. Such approach is used in describing the metamagnetic transitions in  systems with both localised and itinerant spin subsystems ~\cite{Shimizu:1982}. This opportunity for UGe$_2$ is pointed out in ~\cite{Sandeman:2002} in their calculation of density of states which in the expansion of free energy will result in the appearance of M$^8$-term. \\

 \textbf{2. Theoretical approach}\\

 As pointed in the previous section the magnetization in UGe$_2$ is highly anisotropic due to localized 5f-electrons. The a-axis is the easy magnetization axis along which  $M_a$, coming from both itinerant and localised 5f-electrons has the greatest value. The magnetic moment coming from itinerant 5f-electrons is isotropic and, in principle, also the perpendicular part of magnetisation ($M_b$, $M_c$), where b and c are the crystal axes, should be taken into account. In  ~\cite{Shimizu:1982} where metamagnetic transitions are considered in the presence of both localised and itinerant electrons such problem does not appear as the calculations there are made in the presence of external magnetic field which orders the magnetic moments along it.
  As a first step in considering the phase transitions between the paramagnetic phases and FM1 and FM2 at ambient pressure we will not take into account  the transverse magnetisation as small compared to the magnitude of magnetisation along  $a$-axis and will expand the Landau free energy $F(M)$ in $\overrightarrow{M}=(0,0,M_a)$, the subscript $a$ will be omitted hereunder.
 Then $F(M)=f(M)V$ with $V$ - the volume where  the free energy density $f(M)$ is given by the expression:
 \begin{equation}\label{Eq1}
   f=a M^2 +\frac{b}{2}M^4 +\frac{c}{3}M^6+\frac{v}{4}M^8.
 \end{equation}

\begin{figure}[!ht]
\begin{center}
\includegraphics[scale=0.55]{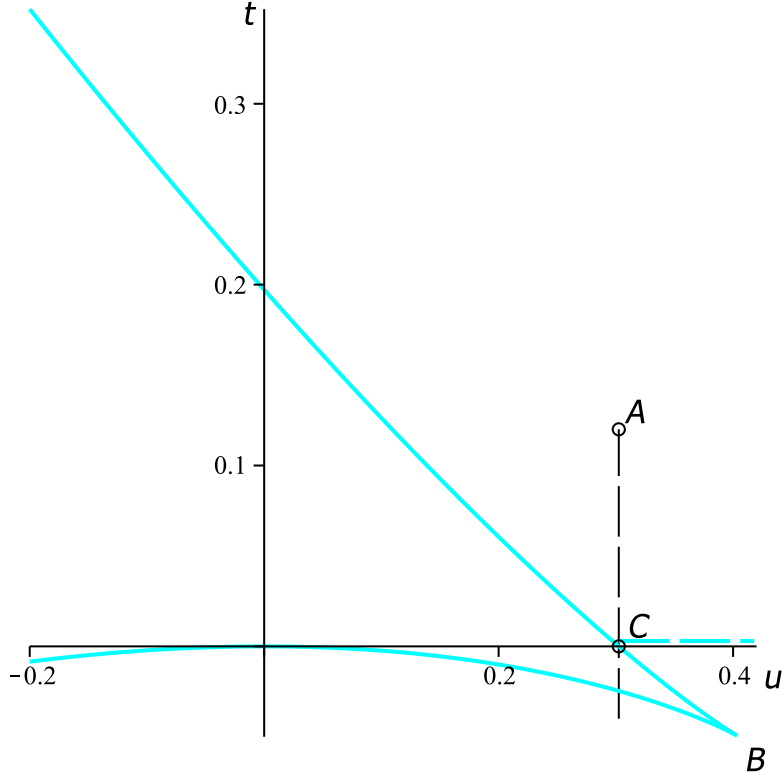}
\end{center}
\caption{The solutions $t_{1,2}$ of quantity Q as function of the parameter $u$. The dashed vertical curve is drawn at $u=\frac{w^2}{4}$ marked by point C in the figure. Point B is at $u=\frac{w^2}{3}$.}
\label{Fig2}
\end{figure}

In the above equation, $a=\alpha (T-T_c)$, where $T_c$ is the critical ferromagnetic temperature and $\alpha$ is a material parameter. The other coefficients in the Landau expansion, namely, $b, u, v$ are considered at this stage not dependant on temperature and external pressure; $b, c$ may be either positive or negative, but $v>0$ in order to ensure converging of $f$.
We will redefine  the order parameter $M$ to make the free energy density dimensionless by introducing:
\begin{equation}\label{Eq2}
  m=v^{1/8}M.
\end{equation}
The dimensionless free energy density of Eq. (\ref{Eq1}) will be:
\begin{equation}\label{Eq3}
  f= tm^2+\frac{u}{4}m^4+\frac{w}{3}m^6+\frac{1}{4}m^8;
\end{equation}
and the coefficients in Eq. (\ref{Eq3}) are related to the initial ones in the following way:
\begin{eqnarray*}
       \nonumber u &=& \frac{b}{v^{1/2}}\\
      \nonumber  w &=& \frac{c}{v^{3/4}}
     \end{eqnarray*}
The reduced temperature is $t =\beta (T/T_c-1)$ with $T_c$ - the Curie temperature, $\beta=\alpha T_c/v^{1/4}$.
The equation of state $(df/dm)$ :
\begin{equation}\label{Eq4}
2m(t+um^2+wm^4+m^6)=0
\end{equation}
has an obvious solution for disordered - paramagnetic phase $m=0$.
If we substitute $x=m^2 \geq 0$, the above equation will become standard 3-rd order algebraic equation whose solutions can be analytically expressed, see, for example~\cite{Abramowitz:1964}. The Eq. (\ref{Eq4}) expressed by new variable $x$ reads:
\begin{equation}\label{Eq5}
t+ux+wx^2+x^3=0.
\end{equation}
The number of real solutions for $x=m^2$ may be determined by the sign the quantity
\begin{equation}\label{Eq6}
Q=\frac{t^2}{4} +\frac{2w}{3}(\frac{2w^2}{9}-u)t+\frac{u^2}{27}(u-\frac{w^2}{4})
\end{equation}
The analysis of $Q(t)$ which is a quadratic equation with respect to $t$ as function of parameters $u,w$ with solutions $t_{1,2}$ given by:
 \begin{equation}\label{Eq7}
 t_{1,2}=\frac{w}{3}(u-\frac{2}{9}w^2)\pm \frac{2}{27}(w^2-3u)^{3/2}
 \end{equation}
also determines the region of existence of non-negative solutions for magnetisation as function of parameters $u$ and $v$ in Landau energy. It is obvious that $t_{1,2}$ are real for $w^2\geq 3u$ and $t_0=w^2/(3u)$ is a special point, for which $t_1=t_2=w^3/27$.
  Q may be written as
  \begin{equation}\label{Eq8}
  Q=(t-t_{1})(t-t_{2}).
  \end{equation}
  Then for $t$ between t$_{1,2}$, Q$<0$ and Eq.(\ref{Eq5}) will have three real solutions; for $t> t_{1}$, and $t< t_{2}$, Q$>0$ and there are one real positive solution and two complex conjugate solutions; for Q$=0$ - two real equal solutions, and one more real solution.\\
 An example of the dependence of $t_{1,2}$ on the parameter $u$ is presented in Fig. \ref{Fig2} for $w =-1.1$.
 
\begin{figure}[!ht]
\begin{center}
\includegraphics[scale=0.55]{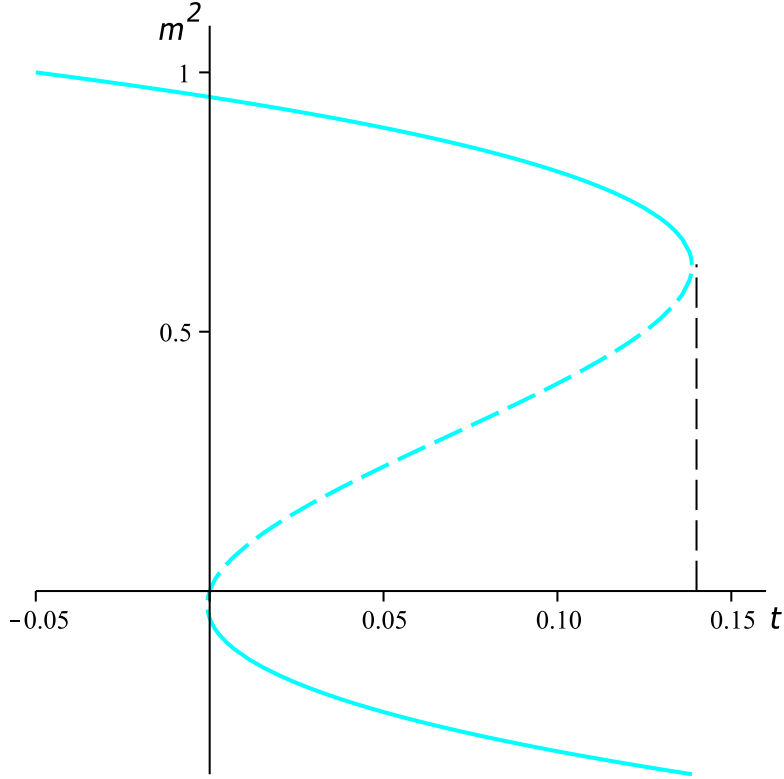}
\end{center}
\caption{Three real solutions of (\ref{Eq5}) for magnetisation $m^2=x$ as function of reduced temperature for $u=-0.05$ and $w=-0.9$. The dashed blue line describes solution $x_3$, which is metastable and the solid blue curve below t-axis - the solution $x_2<0$. }
\label{Fig3}
\end{figure}

 For the analysis of solutions of equation of state~Eq.(
 \ref{Eq5}) it is important to find out the regions where $x=m^2$ is non-negative. This is why  we will present the  solutions of~(\ref{Eq5}) in analytical form by introducing the variables
  \begin{subequations}\label{Eq9}
\begin{align}
s_1&=\left[\frac{w}{6}(u-\frac{2}{9}w^2)+\frac{t}{2}+\sqrt{Q} \right]^{1/3} \\
s_2&=\left[\frac{w}{6}(u-\frac{2}{9}w^2)+\frac{t}{2}-\sqrt{Q} \right]^{1/3}
\end{align}
\end{subequations}
  by which the solutions $x$ can be presented as follows:
   \begin{eqnarray}\label{Eq10}
   x_1&=&s_1+s_2-\frac{w}{3}\\
  \nonumber x_2&=&-\frac{s_1+s_2}{2}-\frac{w}{3} +\imath \frac{\sqrt{3}}{2}(s_1-s_2)\\
  \nonumber x_3&=&-\frac{s_1+s_2}{2}-\frac{w}{3} -\imath \frac{\sqrt{3}}{2}(s_1-s_2)
   \end{eqnarray}
 The calculations show that for $0<u\leq 0$ , $x_1$ and part of $x_3$ are positive, $x_2$ is negative in the region between $t_1$ and $t_2$ and no isostructural transition can take place there. The phase transition from paramagnetic to ferromagnetic phase is of first order as seen from~Fig. \ref{Fig3}, where  the three real solutions for $m^2=x$ as function of reduced temperature $t$ and $u<0$ are shown, also the transition to ferromagnetic phase is shifted to positive values of $t$.
We note that our calculations are made for $w<0$ as the estimate shows that for $w>0$ it will be redundant to expand the Landau energy  up to $M^8$, as no new phases will appear and only some redefining of values of parameters in Landau energy will occur  compared to the expansion of the Landau energy to $M^6$.\\
For $0<u<w^2/4$, $x_1,\; x_3$ are positive and $x_2$ changes sign and passes through zero and the negative part of $x_2$ decreases and is zero at $u=w^2/4$. It is obvious from Fig. (\ref{Fig2}) that in this range of parameter $u$ the phase transition from the disordered to ferromagnetic phase is of first order and with the decrease of temperature an isostructural first-order transition occurs. Even for  $u=w^2/4$ the phase transition from paramagnetic phase to the magnetically ordered is of weak first order succeeded by isostructural transition at temperature lowering. For $w^2/3>u>w^2/4$ two phase transitions occur: second order transition from disordered to low-magnetisation phase and first order isostructural transition to high-magnetisation phase when temperature decreases. This region of parameters $u$ is illustrated in Fig. (\ref{Fig2}) by blue dash line. \\
The stability conditions are given by inequality:
\begin{equation}\label{Eq11}
\frac{d^2f}{dm^2}=2(t+3um^2+5wm^4+7m^6)\geq 0
\end{equation}
and there are several methods to resolve this condition in analytical form: see for example ~\cite{Izyumov:1984}. The problem may be also done numerically by direct substitution of solutions of (\ref{Eq4}) in the above equation and analysing its positiveness. If more than one solution is stable also a comparison between free energies should be made in order to find out which solution in what domain of reduced temperature is an absolute minimum. \\

\textbf{3. Results and discussion}\\

\begin{figure}[!ht]
\begin{center}
\includegraphics[scale=0.55]{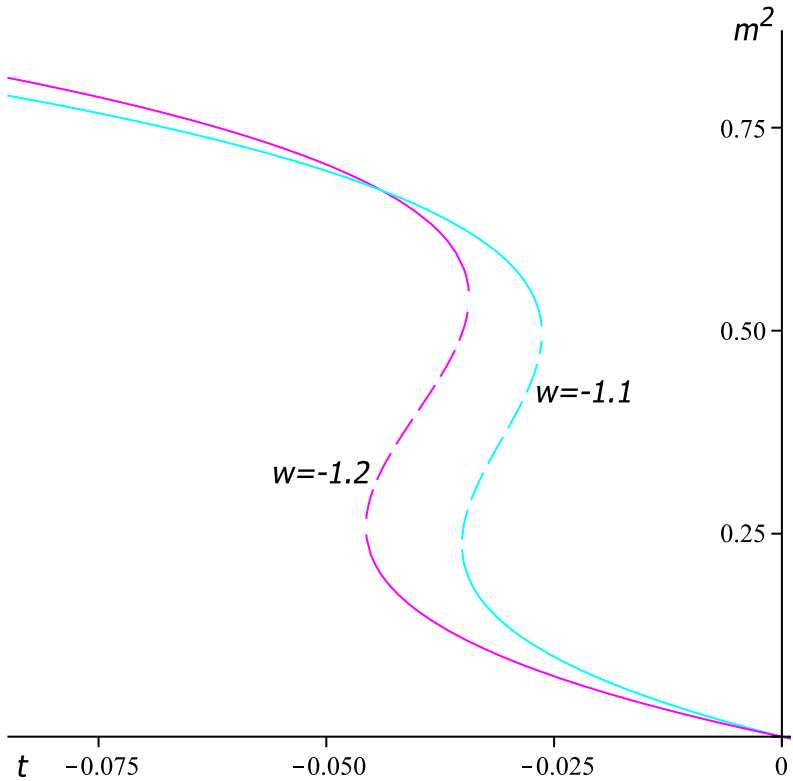}
\end{center}
\caption{The magnetisation $m^2=x$ as function of reduced temperature for different values of $w$; for both curves $u=(\frac{w^2}{4}+\frac{w^2}{3})/2$ for the respective value of $w$. The dashed  lines describes the metastable solution $(x_3)$ of Eq. \ref{Eq10} for magnetisation. }
\label{Fig4}
\end{figure}

The calculations show that the most important region in Fig.~\ref{Fig2} is where $Q\leq 0$, because there is possibility for the appearance of two different phases with same structure but different magnitude of magnetisation as in UGe$_2$. As the experiment shows that the phase transition from paramagnetic to FM1 is of second order we focus our attention on the region of parameter $w^2/4< u <w^2/3$, given by dashed horizontal blue line in Fig.~\ref{Fig2}, where the possibility for second order transition from disordered phase to low-magnetisation phase is present. We illustrate the behaviour of magnetisation in  this region of parameter $u$ in Fig. \ref{Fig4}.
The solution of Eq. (\ref{Eq5}), $x_2$ given by Eq. (\ref{Eq10}) describes the appearance of low magnetisation  from the disordered phase. It passes through the point $t=0$ and exists only for $t<0$; for $t>0$ it is negative. With the decrease of reduced temperature $t$  a first order phase transition occurs to  high magnetisation phase  given by $x_1$ (see, Eq. (\ref{Eq10}). The dashed lines describe the metastable solution $x_3$ (see, Eq. (\ref{Eq10}). The light blue line in the figure shows the magnetisation for $u=w^2/4$ ; point C in Fig. (\ref{Fig2}).\\
As microscopic calculations including the dual nature of 5f-electrons at ambient pressure are absent, neither the magnitude nor the sign and especially the  temperature dependence of parameters $u$ and $w$ can be estimated. We here can only point out the relations between these parameters for the Landau energy expansion up to $M^8$. Both analytical and numerical analysis show that the parameter $w$ determines the temperature width of low-magnetisation phase, namely $w^2/4,w^2/$. The bigger  magnitude of $w$ defines wider temperature interval of low-magnetisation phase and wider choice of the parameter $u$ in front of the $m^4$ term in the Landau expansion Eq. (\ref{Eq3}).
\begin{figure}[!ht]
\begin{center}
\includegraphics[scale=0.55]{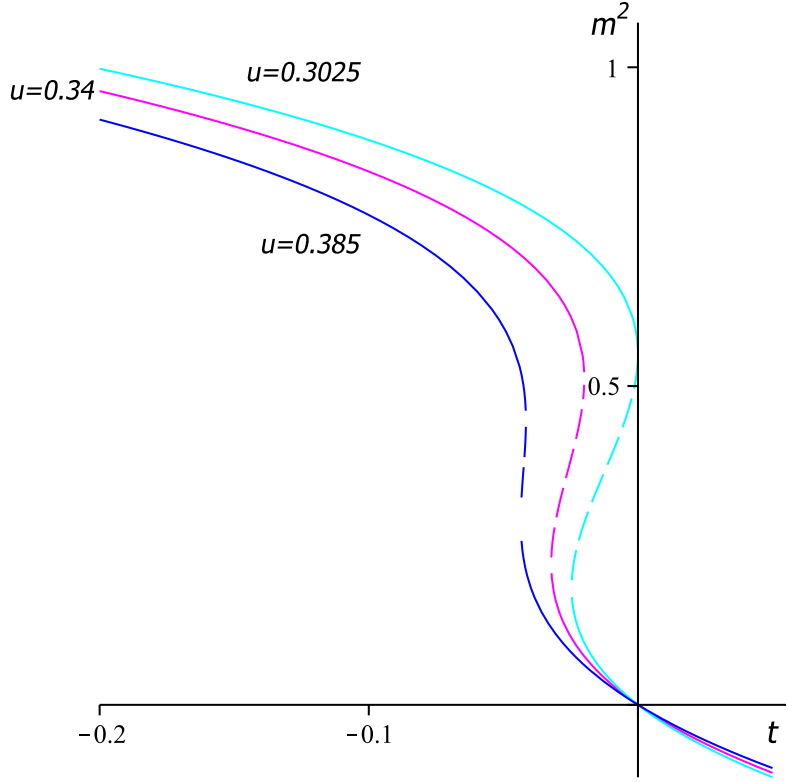}
\end{center}
\caption{The magnetisation $m^2=x$ as function of reduced temperature for different values of $u$ and $w=-1.1$. The dashed  lines describes the metastable solution for magnetisation. }
\label{Fig5}
\end{figure}
It is seen from the figure that the value of parameter $u>0$ defines to some extent the width of low-magnetisation phase with respect to the reduced temperature $t$ for the chosen parameter $w$. As $u$ increases this width increases too. There is strong dependence between the magnitudes of parameters $u$ and $w$, and obviously for the appearance of isostructural transition from low- to high-magnetisation phase $|w|>u>0$. In principle this can be related not directly with the experimental value $(T_x-T_c)/T_c\simeq -0.43$ of UGe$_2$ but in order to make reliable estimate an adequate microscopic calculations should be made for magnetic phase transitions at ambient pressure. Also the fact of working in quite rough approximation, not taking into account both transverse components of magnetisation and dropping the temperature dependence on temperature especially of coefficient $u$ before $M^4$ in Landau energy also limits the possible variations of both$w$ and $u$.\\

The parameter $u>0$ is always smaller in magnitude than $w$ as is seen from Fig. (~\ref{Fig2})in the area of possible isostructural transition, i.e., between  $t_{1,2}$ given by Eq. (\ref{Eq7}). As we are interested in values of $w^2/3 >u>w^2$ where second order phase transition from disordered to low-magnetisation phase followed by isostructural first order transition to high magnetisation phase appears, we will focus there the analysis of the magnetisation as function of parameter $u$ for fixed value of $w$. It is illustrated in Fig.(\ref{Fig5}) for $w=-1.1$, and it is obvious that parameter $u$ influences the width of temperature interval for the existence and stability of low magnetization phase. From the experiment $\Delta T=(T_x-T_c)/T_c \simeq 0.43$. The respective temperature interval expressed by the reduced temperature as $\Delta t = \beta(T_x-T_c)T_c$ with $\beta =\alpha T_c/v^{1/4}$ cannot be directly evaluated and compared with $\Delta T=(T_x-T_c)/T_c$ as   the parameter $\beta$ should be determined by the respective microscopic calculations.\\  
 The line which describes the isostructural transition is determined by making equal the minima of free energy $f$ (Eq. (\ref{Eq3}) of low- and high-magnetisation phases , namely 
 \begin{equation}\label{Eq12}
 f(m_1)=f(m_2),
 \end{equation}
 where 
 $$m_1=\sqrt{x_1}$$ and 
 $$m_2=\sqrt{x_2}$$.
  \begin{figure}[!ht]
\begin{center}
\includegraphics[scale=0.55]{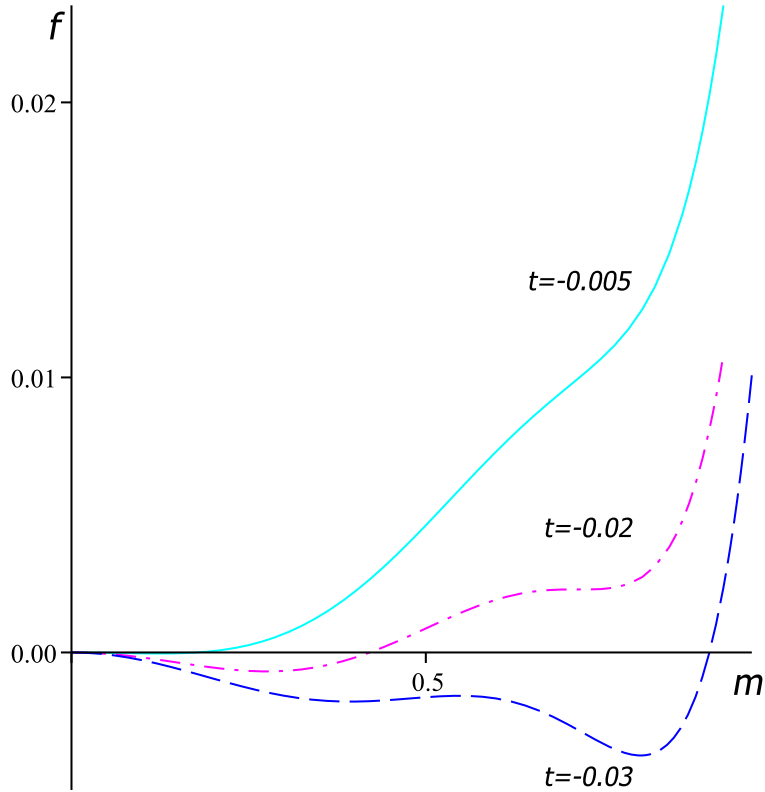}
\end{center}
\caption{Landau free energy as a function of m for $w=-1.1,\; u=0.34$ at different values of reduced temperature$t$}
\label{Fig6}
\end{figure}
Taking into account the equation of state Eq. (\ref{Eq4}) we calculate the line of isostructural phase transition as: 
\begin{equation}\label{Eq13}
t_i=\frac{w}{3}\left(u-\frac{2}{9}w^2\right)
\end{equation}
\begin{figure}[!ht]
\begin{center}
\includegraphics[scale=0.55]{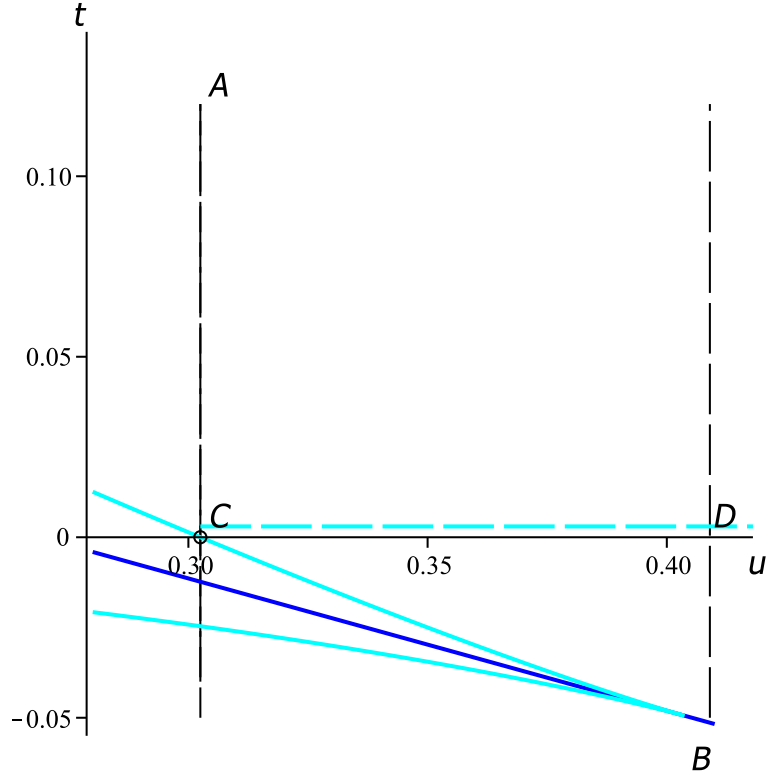}
\end{center}
\caption{The phase diagram for Landau energy expansion up to $M^8$  in coordinates $(u,t)$. The light blue lines define the region of magnetic phases existence and the dark blue line marks the first order phase transition from low-to high-magnetisation phase  }
\label{Fig7}
\end{figure}
The phase transition line from Eq. (\ref{Eq13}) is shown in Fig. (\ref{Fig7}) in the interval $w^2/4<u<w^2/3$. We suppose that  the phenomenological description of magnetic phase transitions at ambient pressure with the help of Landau expansion up to $M^8$ with coefficient in front of $M^4$ falls in the interval between the two black dashed lines passing through points C and D respectively, in the above Figure.
It will be important to include in the above phenomenological study the dependence of parameter $c$ before the $M^4$ -term  on temperature, see Eq. (\ref{Eq1}) , for example of the form $c\sim \gamma (T-T_x)$, in order to describe more realistically the magnetic phase transitions at ambient pressure for UGe$_2$. Further microscopic calculations are needed to shed light on the microscopic mechanism of FM1$\rightarrow$ FM2 phase transition at ambient pressure.\\
A further step is to take into account the transverse components of magnetisation to the lowest order in Landau expansion having in mind the small magnitude of magnetisation , compared to that along the easy magnetisation axis. From experimental data the ratio between transverse and longitudinal magnetisations is $\sim 0.002$, so such approach may be justifed to see what is the effect of transverse magnetisation terms on the order of phase transitions.\\
As far as it is generally accepted that the spin fluctuations are responsible for occurrence of superconductivity of p-type, it will be of great interest also to consider their role within the phenomenological approach.
 The above problems will be the subject of further studies. 

\textbf{4. Concluding remarks}\\
In this paper we propose to apply the phenomenological Landau approach to the description of magnetic phase transitions at ambient pressure far from superconducting transition. The aim is to find a model, by which the occurrence of two different magnetic phases with same structure but different magnetisations can be described. For this aim we propose expansion of free energy up to $M^6$, for which by general theory~cite{Izyumov:1984},~\cite{Toledano:1987} it has been proven that for particular values of Landau parameters, there is a possibility for isostructural transition to occur. 

\textbf{Acknowledgements}\\
This work is supported by Grant KP-06-N38/6 of the Bulgarian National Science Fund.

\end{document}